\documentclass[sigconf]{acmart}

\usepackage{todonotes}

\AtBeginDocument{%
  \providecommand\BibTeX{{%
    \normalfont B\kern-0.5em{\scshape i\kern-0.25em b}\kern-0.8em\TeX}}}

\copyrightyear{2020}
\acmYear{2020}
\setcopyright{acmlicensed}
\acmConference[PDC '20: Vol. 2]{Proceedings of the 16th Participatory Design Conference 2020- Participation(s) Otherwise - Vol. 2}{June 15--20, 2020}{Manizales, Colombia}
\acmBooktitle{Proceedings of the 16th Participatory Design Conference 2020- Participation(s) Otherwise - Vol. 2 (PDC '20: Vol. 2), June 15--20, 2020, Manizales, Colombia}
\acmPrice{15.00}
\acmDOI{10.1145/3384772.3385142}
\acmISBN{978-1-4503-7606-8/20/06}

\begin{document}

\title{p for Power: Need for Agency before Political Participation}
\title{A for Agency: Building Towards \emph{p}olitical Participation}
\title{A Before P: Agency Before Participation}
\title{P for Power: Building Agency towards \emph{p}olitical Participation}
\title{p for \emph{p}olitical: Building Agency For and Beyond Participation}
\title{p for \emph{p}olitical: Participation Without Agency Is Not Enough}
\author{Aakash Gautam}
\authornote{We apologize to Eevi Elisabeth Beck for appropriating the title from \cite{beck2002p}.}
\affiliation{%
  \institution{Virginia Tech, USA}}
\email{aakashg@vt.edu}

\author{Deborah Tatar}
\affiliation{%
  \institution{Virginia Tech, USA}}
\email{dtatar@cs.vt.edu}



\begin{abstract}
Participatory Design's vision of democratic participation assumes participants' feelings of agency in envisioning a collective future.  But this assumption may be leaky when dealing with vulnerable populations.  We reflect on the results of a series of activities aimed at supporting agentic-future-envisionment with a group of sex-trafficking survivors in Nepal. We observed a growing sense among the survivors that they could play a role in bringing about change in their families.  They also became aware of how they could interact with available institutional resources. Reflecting on the observations, we argue that building participant agency on the small and personal interactions is necessary before demanding larger Political participation.
In particular, a value of PD, especially for vulnerable populations, can lie in the process itself if it helps participants position themselves as actors in the larger world. 

\end{abstract}




\maketitle

\section{Introduction}

Over the past three years, we have been working with an anti-trafficking non-governmental organization (NGO) in Nepal and the sex-trafficking survivors supported by the NGO.
We call this NGO ``Survivor Organization'' (SO) as it was established by a group of sex-trafficking survivors and many staff members at different levels of the organization are trafficking survivors. 
A part of our work involves exploring socio-technical interventions to support the survivors achieve what SO calls ``dignified reintegration''. 
Survivors face a myriad of challenges in reintegration including social stigma held against trafficked persons in Nepali society, hurdles in bureaucratic processes such as in obtaining a citizenship certificate, and lack of employment opportunities \cite{simkhada2008life, poudel2009dealing, joshi2001cheli, sharma2015sex, crawford2008sex, richardson2009sexual}.

While over the past three years we have built a relationship with SO, the space is contested and requires careful goal-balancing.  
The first author is a native Nepali familiar with the cultural norms but as a privileged male researcher from the West,  he is an outsider and is seen as an outsider.
Our knowledge of the complexities of the setting is limited, the goals set by SO for the survivors do not necessarily align with the survivors' visions, and the survivors' dependency on SO for support results in them having limited power to negotiate \cite{redacted}. 
Part of our work has been in seeking ways to support the survivor agency with the hope that they feel emboldened to face challenges arising in their reintegration journey.  

The work reported here began with an exercise to hear how the survivors --- henceforth called \emph{sister-survivors} to better reflect their nomenclature --- envisioned their future with respect to themselves, their family, and others in society. 
While we heard varied expressions regarding their future, they all envisioned a better society, one that did not hate trafficking survivors and other marginalized groups. 
However, they did not see themselves playing a role in bringing about the envisioned changes. 
We then conducted a second session around child marriage and human trafficking, the two problems that the sister-survivors said were common in their villages. 
We discussed factors that cause the problem, listed actors involved in the issue, and  identified ways in which they could interact and act along with those actors to mitigate the problem.

We reflect on the discussions that ensued during the sessions. 
The sister-survivors othered the larger institutions and broader society, and saw themselves being distant from the processes involved in bringing change.
Upon narrowing the lens to look at the societal problems close to their home, they could imagine themselves interacting with known actors to play an active role in bringing about change.
Further, imagining their role in attending to personal, day-to-day interactions, that is, small ``p'' \emph{p}olitical engagement, helped in forming more concrete visions for engagement with larger elements of Nepali society, that is, large ``P'' Political engagement.
We argue that a value of participatory design (PD) can be in enabling the participants to realize their agency in day-to-day interactions. 
This, in turn, builds towards interactions with the larger world. 
We add to Beck's argument for political movement in PD \cite{beck2002p} with a call to focus on the personal \emph{p}olitics rather than institutional Politics, especially when working with vulnerable groups.


\section{Background and Literature Review}

Traditionally, PD has focused on future possibilities and alternatives (e.g. \cite{bjorgvinsson2010participatory, kensing1998participatory, kyng2015creating}). 
It grew during an era where the concern was on the impact of technology in the workplace and the recognition for workers (and unions) to regulate new technology \cite{bodker2018participatory}.
The discourse centered around democratic control and Politics, with a focus on the workers' influence on technology and its adaptation, expansion of choices through alternatives, and engagement with the wider network such as worker unions to strengthen democratic ideals (e.g. \cite{nygaard1975trade, bodker1987utopian, kyng1979systems}). 
The assumption was that the participants had agency to act in relationship with one another and the larger structures such as trade unions to promote their interests in both the design of technology and the broader structures beyond the project. 


Modern PD adapted to technology being commonplace \cite{brereton2008new, bodker2018participatory}. 
With it, the focus moved from the Politics to the project-level relationship between designers and stakeholders, and the various tools, techniques, and facilitation involved in realizing it \cite{bodker2018participatory, kyng2010bridging, light2012human}.
But the assumption of participants' agency such as in promoting self-interests through the use of those participatory tools remained (e.g. \cite{halskov2006inspiration, beck2008instant, bratteteig2012spaces, brandt2006designing}).
For example, Beck \cite{beck2002p}, in arguing for a move to a Political PD, calls for people to become ``stroppy'' users --- users who demand change in the larger system by problematizing their day-to-day interactions. 
However, such problematizing requires agency; 
not all participants may feel that they have such agency.

We see here that the sister-survivors feel that they may not have agency to act in relationship to larger institutions or even their own family. 
This problematizes PD's assumption of participants having agency to  envision future possibilities and alternatives, and raises concern on the effectiveness of participation alone in realizing democratic ideals both within the project and in the broader structure. 
Following Beck's call for PD to ``encompass work motivated in political conscience'' \cite{beck2002p}, we present a possible step towards building agency by encouraging participants to attend to the power relations and interactions in day-to-day life i.e. \emph{p}olitics. 




\section{Methodology}

The work reported here builds on an earlier work where we introduced a web application contextualized around crafting to a group of sister-survivors \cite{gautam2020crafting}. 
SO trained all the sister-survivors to create local handicrafts.
The web application was presented as a way of widening future possibilities by selling the handicrafts online.

During the study, the sister-survivors mentioned that they had limited control in their interactions within and outside of SO. 
We developed a document to support them to envision their future with respect to various aspects of their lives, and with it, chart possible pathways to gain some control. 
The first workshop was conducted with a group of nine sister-survivors in January 2019.
We iterated on the document and the activity based on our reflection from the workshop. 
In this paper, we present findings from the second iteration which we conducted in August 2019. 





\subsection{The Sister-Survivors}
Ten sister-survivors who were living at SO's shelter home participated in the study.
They were between the ages of 11 and 20. 
All of them were enrolled in ``morning schools''; they had classes from 6:30 am to 9:30 am. 
Three of them had just started going to school.
Three others were in the seventh grade, the highest among them, but the school moved students up a grade every six months until the sixth grade so this was their fourth year of formal schooling. 

The sister-survivors are vulnerable in a number of ways: they are young, have experienced traumatic ordeals, many have been shunned by their families, and all of them are dependent on SO for support. 
The opportunities made available to them are influenced by the resources available to SO. 
For example, the sister-survivors started going to school because another NGO sponsored the entire cost for SO to send them to school.

\subsection{Future Envisioning Activity}

We began the activity by eliciting values around six aspects of the participants' lives: 
``me'', ``my family'', ``my society'', ``my crafts and skills'', ``my source of income'', and ``me and my technology''.
We discussed their future vision in relation to those aspects at three different stages of their life: (1) when they are about to leave the shelter home, (2) when they feel they are successful, and (3) when they become old. 
In an earlier iteration, the three stages were 1 year, 3 years, and 5 years.
However, those numbers held little meaning to the sister-survivors since they were of different ages, had been at SO for different duration, and had different outlook for reintegration. 

\begin{figure}
\centering
\includegraphics[width=0.48\textwidth]{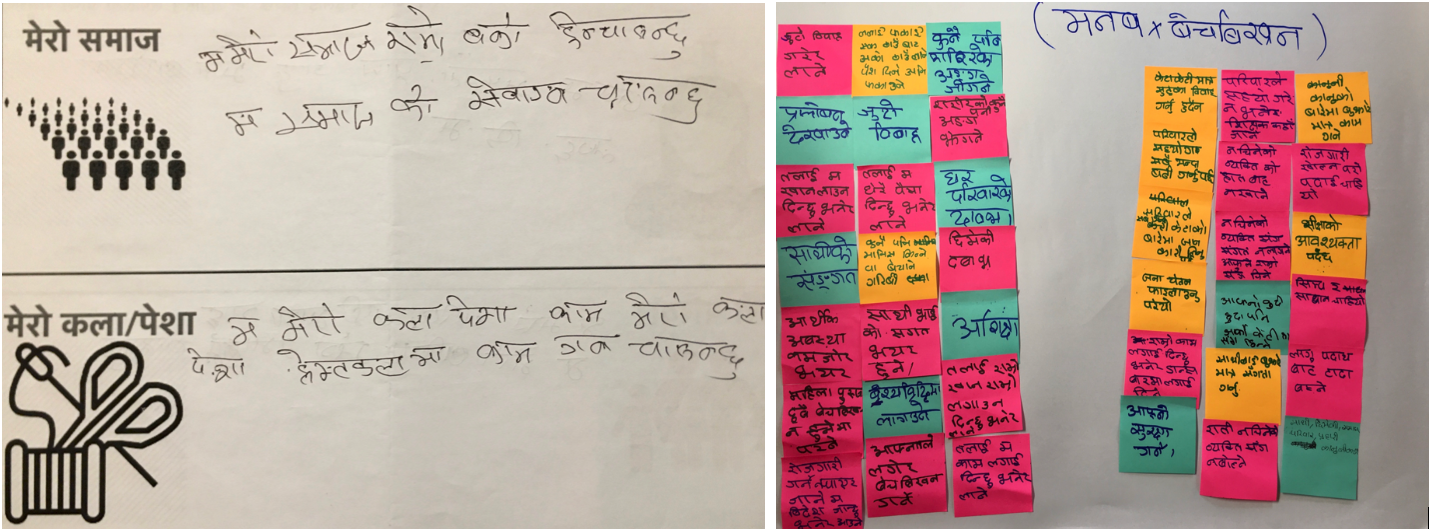}
\caption{Left: S3's vision of her society and her skills when she feels she is successful. Right: Collective response listing the causes, actors, and possible ways for them to be involved against human trafficking.}
\label{fig:sample}
\vspace{-0.3cm}
\end{figure}

The document (see Figure \ref{fig:sample}) created to facilitate the activity was designed considering that many survivors are illiterate \cite{nhrc2018} . 
We had seen SO's walls covered with posters created by survivors using newspaper cutouts, drawings, and texts.  
The document was created such that the sister-survivors could express themselves similarly or simply by speaking about those aspects of their lives.

On the first day, the sister-survivors shared and discussed their views about the future at the time when they are about to leave the shelter home. 
We followed up the next day with a discussion on two societal problems that they had mentioned on the first day: child marriage and human trafficking. 
During this activity, the discussion centered around (1) different factors that lead to the problem, (2) actors who perpetuate and/or can help in mitigating the problem, and (3) ways in which they could act together with the actors to mitigate the problem. 
Eight of the ten sister-survivors participated in this activity; two of them joined us only at the end. 
On the third day, we discussed their vision of the future at the two other stages of their lives. 
Each of the three sessions was around two hours long.


\subsection{Data Collection and Analysis}
At the beginning of each session, we asked for verbal consent from the sister-survivors to record audio and started recording only after obtaining consent from all. 
The first author translated and transcribed the recordings. 
The first round of coding involved closely following the text and summarizing the interactions. 
Following that, the first author merged frequent occurrences into lower-level codes  (e.g. ``Seeing self in raising awareness'').
Following Salda{\~n}a, these codes were discussed with the second author, and in multiple rounds, the codes were merged into higher-level codes \cite{saldana2015coding}.
While the data cover multiple aspects of the sister-survivors' lives, here we focus on their expressions with regards to themselves, their families, and members and institutions in their society.

\vspace{-1.2mm}
\section{Findings}
We observed elements of the sister-survivors' initial, limited agency with respect to their family and institutions. 
As the sessions progressed, they saw possibilities of interacting with other actors and playing an active role in bringing about the changes they imagined.

\vspace{-1.2mm}
\subsection{Limited Agency When Envisioning Future}

\subsubsection{Lack of Agency Within Family}

Most of the sister-survivors felt that they were placed at the fringe of their family's decision-making process.  
They discussed the pressure imposed upon them by family members trying to model others in the community. 
For example, S10 mentioned, ``\textit{They }[family] \textit{will say things like `so and so's son and daughter are now married, you too should marry'}, [that]\textit{ creates stress.}''
Similarly, the influence of societal norm of discriminating against daughters came up frequently as heard in S8's statement, ``\textit{they }[family]\textit{ don't educate daughters. They educate only sons.}''
To this, S9 added that daughters are seen by families as ``\textit{types that leave}'', suggesting their limited influence in their family's decisions.   
Similarly, S6, recounting observations from her village, noted the significance of family pressure on young girls, ``\textit{Everyone in our village runs away and does it} [marries]'' and added that they elope at a young age ``\textit{because the parents try to force them. So, they choose their own and run away instead.}''


The sister-survivors expressed further marginalization in decision-making resulting from the pressure imposed upon their family by members of the society. 
S8, for example, highlighted the way others pressurize their family, ``\textit{They say, `Your daughter has grown up, now get her married' and that puts pressure} [on the family]'' which led S7 to comment, ``\textit{More than the parents, it is those outsiders that put pressure.}''
Similarly, when discussing factors that cause human trafficking, S5 mentioned that a major factor of trafficking is the societal pressure on families to send daughters to work.
These intricate relationships between the influence members of society have over families, and the power family members hold over a girl child's life repeatedly came up leading S5 to comment, ``\textit{Everything seems to lead to that same thing: family and neighbors.}''

\vspace{-0.5mm}
\subsubsection{Distant View of Institutions}

The sister-survivors expressed views of being distant from existing institutions. 
They mentioned that the police could play a role in tackling both child marriage and human trafficking. 
Some, like S1, thought that members of the society and the police could get together to find a resolution to social problems whereas S6 and S9 saw the police's role limited to raise awareness in villages.
But none expressed views in which they or others in the society could leverage the police for help.

Similarly, the sister-survivors wanted the government to help but felt distant from the government and did not see ways to engage with the government or to seek support from it.
S8 mentioned that child marriage is prevalent despite being illegal because ``\textit{the government overlooks these issues in most places.}''
S10 too felt that the ``\textit{government does not pay attention}'' to mitigate child marriage.
Similarly, S2 felt that the government did not take enough responsibility in mitigating human trafficking and expressed, ``\textit{The government should raise public awareness through programs like street plays.}''
This is particularly noteworthy because S2 had conducted street plays for SO and yet did not see herself playing a role in working with institutions to raise awareness.

\subsubsection{Limited Possibilities for Action}
All of them wished for a better society by the time they leave SO 
and were concerned about being accepted as heard in S8's expression, ``\textit{I wish that when I join society, it doesn't look down upon me. May my society not hate me.}''
S4 expressed her wishes more positively, ``\textit{I want my society to look at me with kindness and in a good light. May it love me and may it do good to other survivors like me.}''  
None except S9 mentioned a role they could play in helping move the society towards their vision. 
S9 wrote, ``\textit{I want to be able to say that my society is very nice. I wish nothing bad goes on in my society}'' and added, ``\textit{I also want to help in making my society better}'', but when asked, she did not know what she could do to make her society better.



The lack of agency in influencing members of the society could be heard throughout the earlier discussions.
For example, S10 mentioned the helplessness survivors may feel when they encounter hatred from others in society, 
``\textit{It is hard to mix with society. One may say something today, someone else may say something later on, and someone else may say something, and that will irritate them} [survivors] \textit{and they can't stay there anymore. They may not have any option left.}''
When we asked about possible actions to reduce hate and discrimination, S10 suggested ``\textit{raising awareness such as through street plays}'', an approach that modeled SO's operation but did not see herself doing it. 
Like S10, others too expressed views where they did not see themselves being involved in taking action.

\vspace{-0.5mm}
\subsection{Envisioning \emph{p}olitical Action}

As the sessions progressed, the sister-survivors saw possibilities to make their families stronger, leverage institutional resources, and also act on their own.  
They discussed ways in which they could interact with known actors such as their family members and neighbors, suggesting an increasing move towards \emph{p}olitical action.

\subsubsection{Making Family Stronger}

All the sister-survivors saw a need for families to be strong as heard in S10's expression, ``\textit{Our own family has to be strong, that's the main thing.}''
To this, S8 added, ``\textit{First, mom and dad have to live harmoniously together. That's needed. Then others can't look down upon us. All} [family members] \textit{have to love each other. That probably will help.}''

They discussed ways in which they could help strengthen their families.
Some mentioned the need for themselves to have confidence and strength before supporting their family as heard in S7's plan of action, ``\textit{First, we have to have self-confidence and be strong. That's needed.}''
S2 mentioned the need to become a model example and raise awareness and, like S8, also suggested a need for family cooperation, ``\textit{First of all, we have to be good. Also, dad and mom should listen to one another. And we should raise a bit of awareness among family members and outside} [the family].''

Like S2, others too saw themselves interacting with family and community members to raise awareness. 
They had earlier mentioned that people were unaware of factors that lead to societal problems.
To address this, S8 wanted to raise public awareness by ``\textit{telling them} [family and neighbors] \textit{that they shouldn't do so} [child marriage].''
Such assertive action plans were increasingly formulated as the sister-survivors began charting out various factors and actors involved in the societal problems, and ways in which they would interact with those actors in their day-to-day lives.

\subsubsection{Interacting with Existing Institutions}

The sister-survivors acknowledged that they alone may not be able to raise awareness and bring about change. 
As the second session progressed, they expressed the need to engage with others and leverage external resources to tackle societal problems.
Particularly, they expressed a desire to leverage existing institutional power held by different local actors as heard in S10's plan to ``\textit{bring in the police or NGOs or other people like teachers, who can help, people who can advise families}'' while discussing ways to mitigate child marriage.

Once the sister-survivors identified the police as potential actors who could help, they discussed plans to engage with the police. 
Acknowledging the power of the police to deter community members from perpetuating societal problems, they saw possibilities of getting the police to ``\textit{advise and warn families and neighbors}'' (S6).
Further, they saw possibilities of engaging the police to arrest people, including their own family members, if the people did not heed to advice and warning. 
The sister-survivors also saw the potential of the police as a resource for support.
It could be heard in S8's fallback plan if her family did not support against trafficking, ``\textit{... the family should support} [against trafficker]\textit{ as needed. If the family does not support, then} [I will] \textit{go to the police.}''

However, the sister-survivors were also concerned about the excessive use of police force. 
S10, for example, mentioned, ``\textit{It's not possible to arrest everyone.}''
This led to a discussion on engaging with other institutions, particularly NGOs and the government. 
S10 had earlier expressed that child marriage is prevalent because ``\textit{ the government does not pay attention}''. 
Later, she saw the possibility to engage and draw the government's attention, ``\textit{if we raise voice in unity, the government may pay attention.}''
She later added, ``\textit{there is nothing we can't do, we can do it but the government has to help a little bit}'', suggesting an interdependence of individual and institutions.

\subsubsection{Identifying Possible Actions}

We could hear more assertive statements as the sessions progressed.
S10 noted that while staff members discussed issues related to human trafficking with the sister-survivors, those issues were ``\textit{never talked about in this detail}'', suggesting a limited understanding of what they could do to mitigate the problem. 
At the end of the second session, S10 added that she now knew what she could do, ``\textit{I have learned a bit about what needs to be done like if I go there }[home]\textit{ and see that child marriage is happening, I feel like I can probably do something. I feel I can at least counsel and advice.}''


Raising awareness in villages and raising their voice to make the government pay attention to overlooked problems in their society were the two major action plans discussed during the sessions.  
All the sister-survivors wanted to create videos and share them online and through television to raise awareness, particularly against human trafficking. 
Even though some of them feared being identified as a trafficked person once they showed their face, all of them expressed willingness to show their faces if needed to convey the message and raise awareness.

\section{Concluding Discussion}
In the sister-survivors' accounts of the future, they wanted a supportive, happy family, and an inclusive and caring society but did not see themselves involved in realizing these visions. 
In fact, their expressions highlighted a lack of agency in their lives, in their family's decision-making process, and in leveraging institutional resources. 
In the second session, we narrowed the institutional Politics to \emph{p}olitics, focusing on problems they had seen in their villages. 
Doing this allowed them to discuss how to change their families and other members of society and, importantly, to imagine themselves playing an active role in bringing about the changes they imagined. 
We eventually saw enough agency for them to imagine engaging actively with the broader Nepali society through, for example, creating videos to raise awareness against human trafficking.

We believe that two inter-dependent aspects of the activity were useful. First, focusing on personal issues that were prevalent close to their home helped them identify factors and the actors involved in those issues.
Second, focusing on their relationship with particular known actors  --- rather than abstract institutions --- helped illuminate possibilities for interaction and invited plans for action.  These approaches broke through the ``othering'' that we saw when the sister-survivors tried to think about more abstract institutions. 
While Politics was perceived to be beyond reach, \emph{p}olitics embodied in personal responses through known relationships was not. 

Imagining engaging with others is not the same as actual engagement. 
Moreover, the known -- and many unknown -- social pressures that the sister-survivors face in society during their reintegration may restrict their actual engagement with other actors.  
In this sense, the work we report here is limited.
A more prolonged and deeper engagement is needed.
But envisionment may have value of its own in their lives and the possibility of partaking in \emph{p}olitics can be of value in preparing them to engage with larger elements of Nepali society, that is, in large ``P'' Political engagement. 

Envisioning future possibilities and alternatives, putting forth self (or group) interest, and negotiating positions and control are fundamental tenets of PD.
In fact, Bødker and Kyng contend that PD -- the PD \textit{that matters} -- needs to move from a focus on co-design sessions towards engagement beyond the project to deal with Political issues around democratic control \cite{bodker2018participatory}. 
They posit that ``democratic control is one of the most challenging'' tasks for PD. They are concerned that modern PD lacks Political outcomes \cite{bodker2018participatory}.  
We agree that the outcome of PD has to be Political for PD to support the realization of democratic ideals. 
However, we have to acknowledge that such political stances require participants' agency to engage in the ensuing interactions and actions. 
In some cases, such as in ours, participants may not feel that they have and indeed may not have sufficient agency. 
Thus, PD must involve enhancing the agency of those who may be on the margins. 
More broadly, ``What are the values that matter?'' and ``Who gets to establish those values?'' are questions that are inherent in PD and should be examined through negotiations between designers, participants, \emph{and} the broader community. 
For PD to matter in issues beyond the sessions, we have to begin by being aware of participants' agency or the lack of it, support in building it if needed, helping them to understand more about power and agency, and emphasize  \emph{p}olitical interaction before striving for democratic control.

\begin{acks}
We wish to thank the sister-survivors and SO staff members for their time and support.
\end{acks}

\bibliographystyle{ACM-Reference-Format}
\bibliography{sample}

\end{document}